\documentclass[amsmath,superscriptaddress,10pt,twocolumn]{revtex4-1}
\usepackage{amssymb}
\usepackage{graphicx}
\usepackage{dcolumn}
\usepackage{bm}
\usepackage{graphicx}
\usepackage{epsfig}
\usepackage{epstopdf}
\usepackage{color}
\usepackage{url}
\usepackage{natbib}

\usepackage{ulem}

\def\nn{\~{n}}
\def\bea{\begin{eqnarray}}
\def\eea{\end{eqnarray}}

\begin{document}
\bibliographystyle{plainnat}
\title{DGP Cosmological model with generalized Ricci dark energy}
\author{Yeremy Aguilera}
\altaffiliation{yeremy.aguilera@usach.cl}
\affiliation{Departamento de Matem\'aticas y Ciencia de la Computaci\'on, Universidad de Santiago, Estaci\'on Central, Santiago, Chile.\\}

\author{Arturo Avelino}
\altaffiliation{aavelino@cfa.harvard.edu}
\affiliation{Harvard-Smithsonian Center for Astrophysics, 60 Garden Street, Cambridge, Massachusetts, 02138, USA \\}

\author{Norman Cruz}
\altaffiliation{norman.cruz@usach.cl}
\affiliation{Departamento de F\'\i sica, Facultad de Ciencia, Universidad de Santiago, Casilla
307, Santiago, Chile. \\}

\author{Samuel Lepe}
\altaffiliation{slepe@ucv.cl}
\affiliation{Instituto de F\'\i
sica, Facultad de Ciencias, Pontificia Universidad Cat\'olica de
Valpara\'\i so, Avenida Brasil 2950, Valpara\'\i so, Chile.}

\author{Francisco Pe\~{n}a}
\altaffiliation{francisco.pena@ufrontera.cl}
\affiliation{Departamento de Ciencias F\'isicas, Facultad de Ingenier\'ia y Ciencias, Universidad de La Frontera, Casilla 54-D, Temuco, Chile}
\date{\today}

\date{\today}


\begin{abstract}
The braneworld model proposed by Dvali, Gabadadze and Porrati (DGP) leads to an accelerated universe without cosmological constant or other form of dark energy for the positive branch $(\epsilon =+1)$. For the negative branch $(\epsilon =-1)$ we have investigated the behavior of a model with an holographic Ricci-like dark energy and dark matter, where the IR cutoff takes the form $\alpha H^2 + \beta \dot{H}$, being $H$ the Hubble parameter and $\alpha$, $\beta$ positive constants of the model. 
We perform an analytical study of the model in the late-time dark energy dominated epoch, where we obtain a solution for $r_cH(z)$, where $r_c$ is the leakage scale of gravity into the bulk, and conditions for the negative branch on the holographic parameters $\alpha$ and $\beta$, in order to hold the conditions of weak energy and accelerated universe.
On the other hand, we compare the model versus the late-time cosmological data using the latest type Ia supernova sample of the \textit{Joint Light-curve Analysis} (JLA), in order to constraint the holographic parameters in the negative branch, as well as $r_cH_0$ in the positive branch, where $H_0$ is the Hubble constant.
We find that the model has a good fit to the data and that the most likely values for $(r_cH_0, \alpha, \beta)$ lie in the permitted region found from an analytical solution in a dark energy dominated universe. We give a justification to use holographic cut-off in 4D for the dark energy in the 5 dimensional DGP model. Finally, using the Bayesian Information Criterion we find that this model it is  disfavored compared with the flat $\Lambda$CDM model. 
\end{abstract}

\maketitle

\section{Introduction}

The acceleration in the expansion of the universe during recent
cosmological times, first indicated by supernova
observations~\citep{Perlmutter} and also supported by the
astrophysical data obtained from WMAP~\citep{WMAP} , indicates the existence of a 
fluid with negative pressure, which have been identified as
dark energy due to its unknown nature. In order to explain the nature 
of this dark energy non conventional approaches have advocated 
extra dimensions inspired by string and superstring theories. 
One of these models that have lead to an
accelerated universe without cosmological constant or other form of
dark energy is the braneworld model proposed by Dvali, Gabadadze,
and Porrati (DGP)~\citep{DGP}, ~\citep{Deffayet},~\citep{Deffayet1}
(for reviews, see ~\citep{Koyama} and~\citep{Miao Li}). In a
cosmological scenario, this approach leads to a late-time
acceleration as a result of the gravitational leakage from a
3-dimensional surface (3-brane) to a fifth extra dimension on Hubble
distances.

It is a well known fact that the DGP model has two branches of
solutions: the self-accelerating branch and the normal one. The self
accelerating branch leads to an accelerating universe without
invoking any exotic fluid, but present problems like
ghost~\citep{Koyama1}. Nevertheless, the normal branch requires a
dark energy component to accommodate the current
observations~\citep{Lue},~\citep{Lazkoz}. Extended models of gravity on
the brane with f(R) terms have been investigated to obtain self
acceleration in the normal branch~\citep{Mariam}. Solutions for a DGP
brane-world cosmology with a k-essence field were found
in~\citep{MariamChimento} showing big rip scenarios and asymtotically
de Sitter phase in the future.

In the present work we explore a DGP cosmology in the framework of the holographic
dark energy models~\citep{Cohen}, ~\citep{Hsu}, ~\citep{Li}, which are based on
the holographic principle~\citep{Gonzalez}. This principle has been advocated as a guideline to a complete theory of quantum gravity. A realization of this principle, based on the validity of the effective quantum field theory, was formulated by Cohen et al~\citep{Cohen}, by making the suggestion that the total energy in
a region of size $L$ should not exceed the mass of a black hole of
the same size, which means $ \rho_{\Lambda}\leq L^{-2}M_{p}^{2}$.
The largest $L$ is chosen by saturating this bound so that we
obtain the holographic dark energy (HDE) density
\begin{eqnarray}
\rho_{\Lambda}=3c^{2}M_{p}^{2}L^{-2},\label{holobound}
\end{eqnarray}
where $c$ is a free dimensionless $\mathcal{O}$(1) parameter that can
be determined by observations. Taking $L$ as the Hubble radius $L =
H_{0}^{-1}$ this $\rho_{\Lambda}$ is comparable to the observed dark
energy density, but gives a wrong EoS for the dark energy~\citep{Hsu}.

For higher dimensional space-times, the holographic principle in
cosmological scenarios has been formulated considering the maximal
uncompactified space of the model, i.e. in the bulk, leading to a
crossing of phantom divide for the holographic dark energy, in 5D
two-brane models~\citep{Saridakis}. Other investigation shows that
when IR cut-off is the event horizon the vacuum energy would end up
with a phantom phase with an inevitable Big Rip
singularity~\citep{Wu}.

Recently, a modified holographic dark energy model has been
formulated using the mass of black holes in higher dimensions and
the Hubble scale as IR cutoff~\citep{Gong}. Using the future event
horizon as IR cutoff, it was found in that the EoS of the
holographic dark energy can cross the phantom divide~\citep{Liu}. The
inclusion of a Gauss-Bonnet term in the bulk and an holographic
energy density have been explored in~\citep{Mariam1}, obtaining a
late-time acceleration consistent with observations. In the same
approach, but using a Ricci-like dark energy, scenarios free of future
singularities were found in~\citep{Mariam2}.

Our aim in this work is to investigate a DGP model of a flat
universe filled with an holographic Ricci-like dark energy~\citep{CGao}
and dark matter. This holographic energy density takes the form~\citep{Oliveros}
\begin{equation}  \label{holodensity}
\rho_{\mathrm{h}} = (3/8\pi G) (\alpha H^2 + \beta \dot{H}),
\end{equation}
where $\alpha$ and $\beta$ are positive constants. This type of
holographic dark energy works fairly well in fitting the
observational data. Nevertheless, a global fitting on the parameters
of this model using a combined cosmic observations from type Ia
supernovae, baryon acoustic oscillations, Cosmic Microwave
Background and the observational Hubble data do not favor the
holographic Ricci dark energy model over the $\Lambda$CDM
model~\citep{Yuting}. For the far future, the EoS behaves like a
quintom model, crossing the phantom barrier~\citep{Feng},
~\citep{Samuel}. The statefinder diagnostic of this model, in the
framework of general relativity, indicates that interactions in the
dark sector are favored~\citep{Yu}.  It was found that without giving
a priori some specific model for the interaction function, this can
experience a change of sign during the cosmic
evolution~\citep{Samuel1}.

In the case of a DGP model, besides the holographic parameters, 
there exists the parameter $r_c H_0$, where $r_c$ is the
characteristic scale of the DGP model given by
\begin{eqnarray}\label{eq2}
r_{c}=\frac{1}{2}\frac{M_{\left( 4\right) }^{2}}{M_{\left( 5\right)
}^{3}}, 
\end{eqnarray}
which sets a length beyond which gravity starts to leak out into the
bulk. $M_{\left( n\right)}$ is the n-dimensional Planck mass.\\

In this work we are interested in constraining the holographic parameters, $\alpha$, $\beta$ and $r_c H_0$ and make comparison with the $\Lambda$CDM model, using the Bayesian Information Criterion. 
This allow us to stablish what model is most favored by cosmological observations and the suitability of an holographic Ricci-like dark energy in the DGP framework.\\
Our paper is organized as follows. In section II we discuss the DGP model for a flat universe filled with a fluid obeying a barotropic EoS. Constraints for the parameter $r_c H$ are obtained. In section III we study analytically the solution of 
the differential equation given by the model assuming a late-time evolution, where dominates the density $\rho_{\mathrm{h}}$, the solution comes with restrictions from the weak energy condition WEC, the accelerated late-time expansion and the positivity of $H$. In section IV we work in the late-time phase universe and solve numerically a differential equation for $\displaystyle %
E=H/H_{0}$, where $H$ is the Hubble parameter and show a table with the
best estimates for the holographic parameters and figures of the confidence
regions which was obtained for some variables for each branch.
In section V we explain the main calculation to use the SNIa data set and
the Hubble parameter for different redshifts. 
In section VI we give arguments in order to justify the used holographic cut-off in 4D for the dark energy in the DGP model (5D). In section VII we discuss our results obtained for the different branches of the DGP model 
and compare it with the $\Lambda$CDM using the Bayesian Information Criterion.


\section{DGP Model}

For an homogeneous and isotropic universe described by the FLRW metric the
field equation is given by~\citep{Deffayet}, \citep{Deffayet1} (with 8$\pi
G=c=1$)
\begin{equation}
\left( H^{2}-\frac{\epsilon }{r_{c}}\sqrt{H^{2}+\frac{k}{a^{2}}}\right)
=\rho -\frac{3k}{a^{2}}, \label{FLRW}
\end{equation}
where $a$ is the cosmic scale factor, $\rho $ is the total cosmic
fluid energy density on the brane. The parameter $\epsilon =\pm 1$
represents the two branches of the DGP model. It is well known that
the solution with $\epsilon =+1$ represent the self-accelerating
branch, since even without dark energy the expansion of the universe
accelerates. For late-times the Hubble parameter approaches a
constant, $H=1/r_{c}$. In previous investigations, $\epsilon =-1$
has been named the normal branch, where the acceleration only appears
if a dark energy component is included. By considering a flat universe
as it is suggested by the Planck results \citep{PlanckCollaboration}, the Eq.(\ref{FLRW}) becomes
\begin{equation}
H^{2}-\epsilon \frac{H}{r_{c}}=\rho ,  \label{FLRWnullc}
\end{equation}
and the weak energy condition (WEC) implies $r_{c}H\geq \epsilon $. If
cosmic fluid satisfies a barotropic equation of state $p=\omega \rho $, the
conservation equation is given by
\begin{equation}
\dot{\rho}+3H\left( 1+\omega \right) \rho =0. \label{conservationeq}
\end{equation}
From Eqs.~(\ref{FLRWnullc}) and (\ref{conservationeq}), we obtain an
expression for the equation of state parameter, $\omega$, in terms of the Hubble parameter which is given by
\begin{equation}
1+\omega =-\frac{1}{3}\left( \frac{2-\epsilon \left( r_{c}H\right) ^{-1}}{
1-\epsilon \left( r_{c}H\right) ^{-1}}\right) \frac{\dot{H}}{H^{2}}.\label{stateeq}
\end{equation}
According to Eq.~(\ref{stateeq}), $1+\omega <0$ implies that $\dot{H}%
>0$, since $\left( 2-\epsilon \left( r_{c}H\right) ^{-1}\right) /\left(
1-\epsilon \left( r_{c}H\right) ^{-1}\right) >0$ is held for both cases $\left(
r_{c}H\right) ^{-1}\lessgtr 1$. Besides, from this condition we have that WEC implies $r_cH>\epsilon$.\\
For the $\epsilon=+1$ we notice that the leakage scale must be restricted to $r_c H_0>1$. 
For the $\epsilon=-1$ we have $r_cH_0>-1$, wich not implies any further constraint upon $r_cH_0$, 
since we are in the expanding phase with $H_0>0$. 

		\section{ Dark Energy Domination Phase}
		\label{SectionDarkEnergyDominationPhase}

In this section we consider the late-time behavior of our model in the normal branch $\epsilon=-1$, where the holographic dark energy density $\rho_{\mathrm{h}}$ dominates and the matter density $\rho_m$ can be neglected. In this case an analytical solution can be obtained solving Eq.(\ref{FLRWnullc}) for $\rho_{\mathrm{h}}$ given by Eq.(\ref{holodensity}). We obtain the following expression for $r_cH(z)$
\begin{eqnarray}\label{eq9}
r_c H(z)=\frac{1}{\alpha-1}+\left[r_c H_0-\frac{1}{\alpha-1}\right]\left(1+z\right)^{\frac{\alpha-1}{\beta}},
\end{eqnarray}
for $\alpha \neq 1$ and $\beta \neq 0$. We notice from this equation that  $\alpha>1$ is required, to ensure the positivity  of $r_cH(z)$ for an expanding universe. The solution for the scale factor yields
\begin{eqnarray}
a(t)=a_{0}\left[(\alpha-1)r_c H_0\left(e^{\left(\frac{1}{\beta r_c}(t-t_0)\right)}-1\right)+1\right]^{\frac{\beta}{\alpha-1}},
\end{eqnarray}
where the initial condition is $a(t=t_0)=a_0$, and $t_0$ is the present time. Notice that $a(t)$ is well behaved because the exponent $\beta /(\alpha-1)$ is always positive. 
From equations (\ref{holodensity}), (\ref{FLRWnullc}) and the expression for the acceleration $\frac{\ddot{a}}{a}=\dot{H}+H^2$ we have
\begin{eqnarray}
\frac{\ddot{a}}{a}=\frac{H^{2}}{\beta }\left( 1-\alpha +\beta +\frac{1}{r_{c}H}\right),
\end{eqnarray}
we can obtain the conditions upon the parameter $\alpha$ and $\beta$
in order to have an accelerated late-time expansion. These
conditions also ensure that WEC still holds. This late-time expansion behaves like a de Sitter phase.
\begin{table}[h!]
\centering
\begin{tabular}{c|c|c}
\multicolumn{3}{c}{\textbf{Conditions for dark energy domination phase}} \\ \hline\hline
Branch($\epsilon$) 	& $\alpha$ & $\beta$ \\ \hline 
$-1$ & $\alpha>1$ & $\beta>-\left(1-\alpha+\frac{1}{r_c H_0}\right)$
\\ \hline\hline
\end{tabular}
\label{TableCondition}
\end{table}


\section{ The holographic dark energy and matter component}
\label{SectionHubbleParameter}


For a spatially flat FRW universe composed by the holographic dark
energy
as well as a matter component (dark and baryon matter), the Friedmann equation (%
\ref{FLRWnullc}) in the DGP cosmology has the form (with units)
\begin{equation}  \label{Eq-Hubble1}
H^2 - \epsilon \frac{H}{r_c} = \frac{8 \pi G}{3} (\rho_{\mathrm{h}} + \rho_{%
\mathrm{m}}),
\end{equation}
The pressureless matter scales in the usual way as,  $\rho_{\mathrm{m}} = \rho_{\mathrm{m 0}%
} a^{-3}$, where $\rho_{\mathrm{m 0}}$ is the present-day value of the
matter density in the Universe. Inserting the expression (\ref{holodensity}) 
for $\rho_{\mathrm{h}}$ and $\rho_{\mathrm{m}} = \rho_{\rm m0} a^{-3}$ at
Eq. (\ref{Eq-Hubble1}), and reorganizing terms, we have
\begin{equation}  \label{Eq-Hubble2}
\beta \dot{H} - H^2(1-\alpha) + \epsilon \frac{H}{r_c} + \left(\frac{8 \pi G%
}{3}\right) \frac{\rho_{\mathrm{m0}}}{a^3} =0.
\end{equation}

We change the derivative of $H$ with respect to time to the scale factor as $%
\dot{H} = (dH/da)\dot{a} = (dH/da) a H$, then Eq. (\ref{Eq-Hubble2}) becomes
\begin{equation}  \label{Eq-HubbleODE-ScaleFactor}
\beta a \frac{d H(a)}{da} - (1-\alpha)H(a) + \frac{\epsilon}{r_c} + \left(%
\frac{8 \pi G}{3}\right) \frac{\rho_{\mathrm{m0}}}{a^3 H(a)} =0.
\end{equation}

Dividing the Eq. (\ref{Eq-HubbleODE-ScaleFactor}) by the Hubble constant $%
H_{0}$, defining the parameter density $\Omega _{\mathrm{m0}}\equiv \rho _{%
\mathrm{m0}}/\rho _{\mathrm{crit}}^{0}$ where $\rho _{\mathrm{crit}%
}^{0}\equiv 3H_{0}^{2}/(8\pi G)$, changing of variable from the scale factor
to the redshift, and defining the dimensionless Hubble parameter as $E\equiv
H/H_{0}$ , the differential equation (\ref{Eq-HubbleODE-ScaleFactor})
becomes
\begin{equation}
\beta (1+z)\frac{dE(z)}{dz}+(1-\alpha )E(z)-\frac{\epsilon }{r_{c}H_{0}}%
-\Omega _{\mathrm{m0}}\frac{(1+z)^{3}}{E(z)}=0.
\label{Eq-HubbleODE-Redshift}
\end{equation}
We solve numerically this differential equation with the initial condition $E(z=0)=1$, and for both branches, $\epsilon = \pm 1$. 
The values of $( r_c H_0, \alpha, \beta, \Omega_{\rm m0})$ are estimated and constrained using the cosmological observations of type Ia Supernovae as described in the next section.

		\section{Cosmological constraints}
		\label{SectionObservationalConstraints}

We test the viability of the model and constrain its free parameters $%
(r_c H_0, \alpha, \beta, \Omega_{m0})$  by using the  \textit{Joint Light-curve Analysis} (JLA) sample of type Ia Supernovae (SNe Ia) of Betoule et al. 2014 \citep{Betoule2014},  composed by 740 SNe that comes from 9 different surveys.  
We compute the best-fit values and confidence intervals by sampling the parameter space using the Affine Invariant Markov Chain Monte Carlo (MCMC) of Goodman et al. \citep{AffineInvariantMCMC},  implemented in the \textit{emcee} code \citep{emcee}.


The definition of luminosity distance $d_{L}$ in a flat FRW cosmology is given as
\begin{equation}
d_L(z,\vec{p},H_0)=\frac{c(1+z)}{H_0} \int_0^z\frac{dz^{\prime }}{%
E(z^{\prime }, \vec{p})},  \label{luminosity_distance1}
\end{equation}
\noindent where $E(z,\vec{p})$ is given by the numerical
solution to the differential equation (\ref{Eq-HubbleODE-Redshift}), \textquotedblleft$c$\textquotedblright
\ is the speed of light given in units of km/sec and $\vec{p}$ is the vector of parameters, i.e., $\vec{p}=(r_cH_0, \alpha, \beta, \Omega_{m0})$. The theoretical distance modulus is defined as
\begin{equation}
\mu^{\mathrm{t}}(z_k,\vec{p}) =  5\log_{10}\left[\frac{d_L(z,\vec{p})}{\mathrm{Mpc}} \right] +25,
\label{distanceModuli}
\end{equation}
\noindent where the superscript `t' stands for the \textit{theoretical} prediction of the distance modulus for a supernova at a redshift $z_k$.

On the other hand, the observed distance modulus for each supernova can be computed  by modeling their intrinsic variability observed in their light-curves \citep{TrippEquation}, as
\begin{equation}
\mu = m^*_B-(M_B - \alpha_{lc} \times X_1 + \beta_{lc} \times C)
\end{equation}
where $M_B$ is the absolute magnitude of the SNe in rest-frame $B$ band. This parameter, together with $(\alpha_{lc}, \beta_{lc})$ that characterize global properties of the light-curves of the SNe, are nuisance parameters that have to be computed and marginalized simultaneously with the cosmological parameters of interest.

On the other hand,  $m^*_B$, $X_1$, $C$ are the observed peak magnitude in rest-frame $B$ band, the stretch and color parameters for each SN respectively. They capture the intrinsic variability in the luminosity of the SNe. Their central values as well as the covariance matrices that account for all  known sources of systematic uncertainties as well as the statistical uncertainties, are publicly available at \url{http://supernovae.in2p3.fr/sdss_snls_jla/ReadMe.html}.

Following  \citep{Betoule2014}, the distance modulus for all the SNe are reorganized in a vector of $n=740$ entries given as
\begin{equation}
\boldsymbol{\mu} = \mathbf{A} \boldsymbol{\eta} - \mathbf{M_B}
\end{equation}
where the  $n$-dimensional vector $\boldsymbol{\eta}$ and the $n \times n$ matrix $\mathbf{A}$ are given as
\begin{align}
\boldsymbol{\eta} &= \left(  (m^*_{B, 1}, X_{1, 1}, C_1) , ...,  (m^*_{B, n}, X_{1, n}, C_n) \right) \\
\mathbf{A} &= \mathbf{A}_0 + \alpha \mathbf{A}_1 - \beta \mathbf{A}_2, \, \text{with} \, (\mathbf{A}_k)_{i,j} = \delta_{3i, i+k}
\end{align}

With this, the $\chi^2$ function to be minimized to compute the best-fit values and confidence intervals of the cosmological parameters has the form
\begin{equation}
\chi^2(\vec{p}, \vec{p}_{lc}) = \left( \boldsymbol{\mu}(\vec{p}_{lc}) - \boldsymbol{\mu}^{\rm  t}(z, \vec{p}) \right)^{\rm T}  \mathbf{C}^{-1} \left( \boldsymbol{\mu}(\vec{p}_{lc}) - \boldsymbol{\mu}^{\rm  t}(z, \vec{p}) \right) 
\label{ChiSquareDefinition}
\end{equation}
where $\vec{p}_{lc} = (M_B, \alpha_{lc}, \beta_{lc})$ and $\mathbf{C}^{-1}$ is the inverse of the total covariance matrix reported in \citep{Betoule2014} that encapsules all the known systematic and statistical errors. For a detailed discussion on  $\mathbf{C}$ of the JLA sample see Betoule  et al. 2014 \citep{Betoule2014}. 
We consider also the fixed fiducial value of $H_0 = 70$ km s$^{-1}$ Mpc$^{-1}$.
%

\begin{widetext}

\begin{figure}
\begin{center}
\includegraphics[width=16cm]{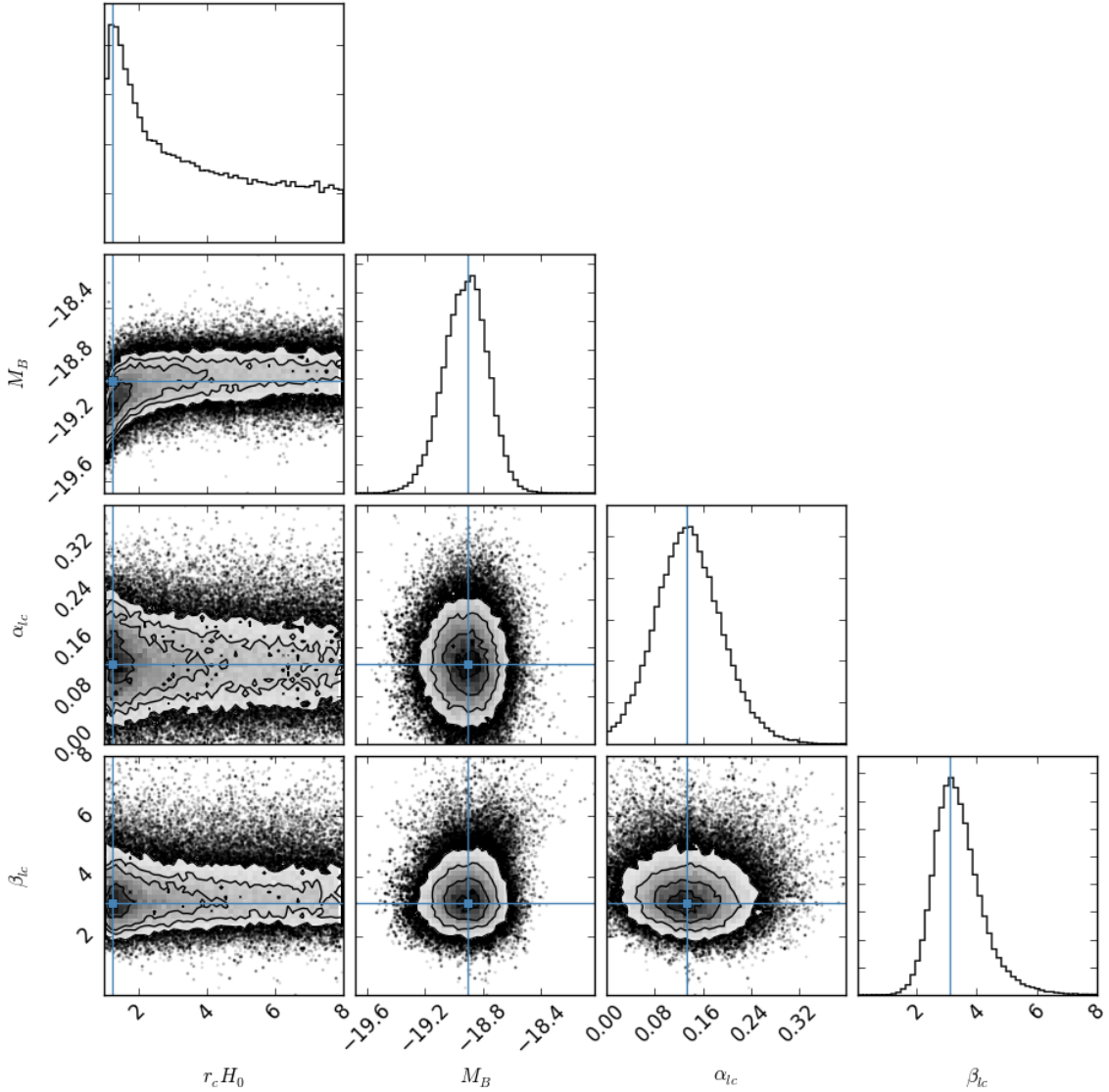}
\caption{Joint and marginalized constraints on the leakage scale $r_c H_0$ of the self-accelerated branch ($\epsilon = +1$) of the DGP model in a spatially flat Universe. It is also shown the global light-curve parameters $(M_B, \alpha_{\rm lc}, \beta_{\rm lc})$ of the JLA sample that were computed simultaneously with $r_c H_0$.
The credible regions correspond to $1\sigma (68.3\%),  2 \sigma (95.5\%)$ and $3\sigma (99.7\%)$ of confidence level (CL). 
We assumed flat priors for all the parameters. We set the physical limit of $r_c H_0>1$.
The individual best-fit values are shown in Table \ref{TableBestEstimated}.}
\label{PlotCRrcHo}
\end{center}
\end{figure}

\end{widetext}

		\subsection{Self-accelerated branch: $\epsilon=+1$}
		\label{SectionEpsilonPositive}

For the positive branch, $\epsilon=+1$, we consider the DGP brane filled with the baryon and dark matter component only, $\Omega_{\rm m0}$. We neglect the dark energy density because this branch is already accelerated. In this case, the only free parameter is the leakage scale $r_cH_0$, given that we set $\alpha = 0, \beta =0$. In this case, the matter density $\Omega_{\rm m0}$ is related to the leakage scale as
\begin{equation}\label{EqRelationOm-rcHoForEPositive}
\Omega_{\rm m0} = 1-(1/r_cH_0).
\end{equation}
This constraint comes directly from the differential equation (\ref{Eq-HubbleODE-Redshift}) for this case.

The marginalized best-fit value for the leakage scale $r_cH_0$ is shown in Table \ref{TableBestEstimated}, item (i), and the Fig. \ref{PlotCRrcHo} shows the joint credible regions for combinations of $(r_cH_0, M_B, \alpha_{\rm lc}, \beta_{\rm lc})$ in pairs, as well as the marginalized probability density functions (PDF) for each parameter.
The relevant result for the leakage scale is that it consistent to have values in the physical region  $r_c H_0 >1$ (see Fig. \ref{PlotCRrcHo}).
Marginalizing over the other parameter, the best-fit value is $r_c H_0 = 1.3^{+3.8}_{-0.3}$.

		\subsection{Normal Branch: $\epsilon=-1$}

For this branch we consider a universe filled with an holographic Ricci-like dark energy and dark matter. We solve numerically the differential equation (\ref{Eq-HubbleODE-Redshift}) with the initial condition $E(z=0)=1$. 
In this case, the free parameters to be estimated are $(r_cH_0, \alpha, \beta, \Omega_{m0})$ together with the light-curve parameters $( M_B, \alpha_{\rm lc}, \beta_{\rm lc})$.

\begin{widetext}

\begin{figure}
\begin{center}
\includegraphics[width=17cm]{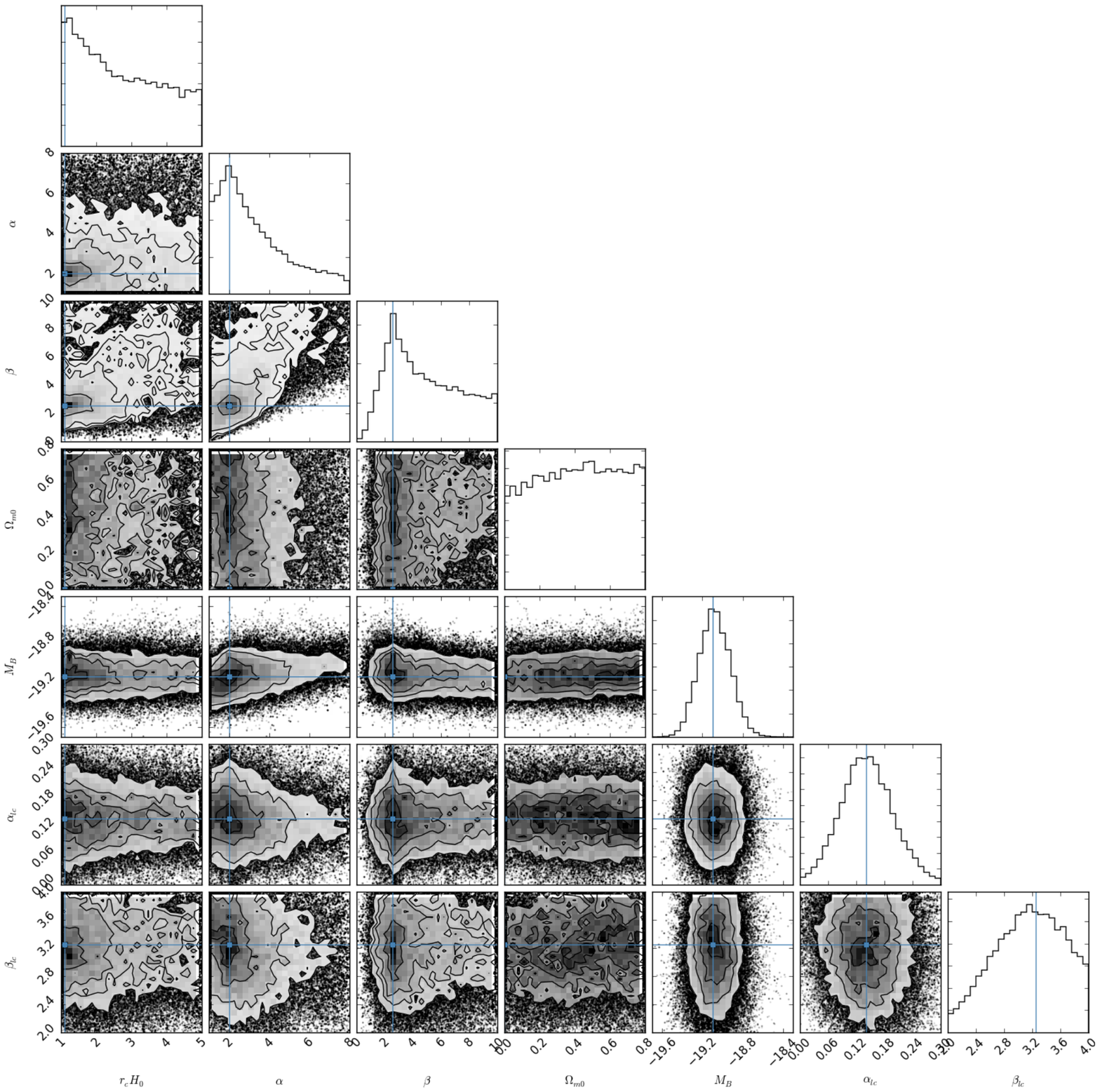}
\caption{Joint and marginalized constraints on the leakage scale $r_c H_0$ of the \textit{non} self-accelerated branch ($\epsilon = -1$) of a DGP model in a spatially flat Universe, filled with a matter component (dark and baryonic matter), $\Omega_{\rm m0}$,  and a Ricci-like holographic dark energy of the form $\rho_{\mathrm{h}} = (3/8\pi G) (\alpha H^2 + \beta \dot{H})$. 
It is shown the marginalized constrains on $(r_c H_0, \alpha, \beta, \Omega_{\rm m0})$ as well as for the global light-curve parameters $(M_B, \alpha_{\rm lc}, \beta_{\rm lc})$ of the JLA sample that were computed simultaneously.
The credible regions correspond to $1\sigma (68.3\%),  2 \sigma (95.5\%)$ and $3\sigma (99.7\%)$ of confidence level (CL). 
We assumed flat priors for all the parameters in the physical intervals:  $0<\Omega_{\rm m} < 1$, $ r_c H_0>1$, $\alpha > 1$,  $\beta > 0$.
The individual best-fit values are shown in Table \ref{TableBestEstimated}.}
\label{PlotCR-All-EpsilonPositive}
\end{center}
\end{figure}

\end{widetext}

Fig. \ref{PlotCR-All-EpsilonPositive} shows the joint credible regions in pairs of parameters as well as their marginalized PDFs. 
We find that the holographic best-fit parameters are $\alpha=2.1^{+3.4}_{-1.1}$, $\beta=2.45_{-1.4}^{+5.5}$ and $r_cH_0=1.2^{+3.5}_{-0.2}$ well compatible with the analytical constraints on the model described in Section II. 

For these parameters we notice that $r_cH_0 = 1.2$, $\alpha=2.1>1$ and $\beta=2.45>-\left(1-\alpha+\frac{1}{r_c H_0}\right)=0.267$, which is in agreement with the constraint derived in Section III for an holographic dark energy domination phase. This values ensure WEC and accelerated expansion.
For $\Omega_{\rm m0}$ we find that the data does not impose tight constraints on it, so that any value in the range $0<\Omega_{\rm m0}<1$ is equally likely according to the data.

\begin{table*}[tbp]
\centering
\begin{tabular}{c c | c c c c  c c  c | c  c}
\multicolumn{10}{c}{\textbf{Best estimates}} \\ \hline\hline 
& Model & $r_c H_0$  &  $\alpha$ & $\beta$ & $\Omega_{\rm m0}$ & $M_B$ &$\alpha_{lc}$ & $\beta_{lc}$ &  $\chi^2_{\mathrm{min}} $  & BIC \\[0.1cm] \hline 

(i) &  DGP $(\epsilon = +1)$ only & $1.3^{+3.8}_{-0.3}$ & $-$ & $-$ & $0.23^{+0.57}_{-0.23}$ & $-18.9^{+0.14}_{-0.15}$ & $0.13^{+0.06}_{-0.05}$ & $3.3^{+0.84}_{-0.65}$ &  684.4 &  710.9 \\

(ii) & DGP $(\epsilon = -1) + \rho_h$  & $1.2^{+3.5}_{-0.2}$ & $2.1^{+3.4}_{-1.1}$ & $2.45_{-1.4}^{+5.5}$ & $0.46^{+0.54}_{-0.46}$ & $-19.08^{+0.16}_{-0.15}$  & $0.14^{+0.056}_{-0.054}$  & $3.17^{+0.5}_{0.55}$  &  683.3 & 729.5 \\[1mm]  \hline
(iii) & flat $\Lambda$CDM & $-$ & $-$ & $-$  &  $0.29 \pm 0.03$  &  $-19.05 \pm 0.02$  & $0.14 \pm 0.01$  & $3.1 \pm 0.01$ &  683.0  & 709.4    \\

\hline\hline
\end{tabular}%
\caption{Marginal best-fit values for the model parameters $(r_cH_0, \alpha, \beta, \Omega_{\rm m0})$ as well as the light-curve parameters $(M_B, \alpha_{lc}, \beta_{lc})$ of type Ia  supernovas of the JLA sample \citep{Betoule2014} and computed together with the cosmological parameters.
The first row shows the best-fit values for the leakage scale $r_cH_0$ for the self-accelerated branch $(\epsilon = +1)$ of a DGP brane model composed only of baryon and dark matter, $\Omega_{m0}$, at late times. 
In this case, $\Omega_{m0}$ is found from the constraint equation $\Omega_{\rm m0} = 1-(1/r_cH_0)$ (see section \ref{SectionEpsilonPositive} for details).
The second row shows the best-fits for $(r_cH_0, \alpha, \beta, \Omega_{\rm m0})$ on the normal branch of a DGP brane model composed by $\Omega_{m0}$ and a Ricci-like holographic dark energy of the form $\rho_{\mathrm{h}} = (3/8\pi G) (\alpha H^2 + \beta \dot{H})$, at late times.
In (iii) it is also shown  the best-fit values for the flat $\Lambda$CDM model in order to compare the results.
We use the Bayesian Information Criterion (BIC) in order to assess the best model to fit the data, between the holographic DGP and the flat $\Lambda$CDM model.
The model favored by the observations compared to the other corresponds to the one with the smallest value of BIC. In general, a difference of 2 in BIC between two models is considered as an evidence against the model with the higher BIC, and a difference of 6 is a strong evidence.
We find that despite the good fit to data by the holographic DGP model, it is  disfavored compared with the $\Lambda$CDM model using the BIC criterion, as shown in the last column. }
\label{TableBestEstimated}
\end{table*}

	\section{Remarks for the holographic dark energy in higher dimensional gravity}

Let us discuss with some detail the results of the studies on the IR cutoff for holographic dark energy models in the framework of higher dimensional gravity. In~\citep{Gong} was considered the mass of the Schwarzschild black hole in $N+1$ dimensional space-time and then using consistently the formulation of Cohen et al~\citep{Cohen}. The direct use of the $N+1$ dimensional solution for a spherically symmetric static matter source can be well justified taking into account that the Schwarzschild radius for the holographic considerations is $~H^{-1}$. In this case the Vainshtein radius, $(r_{c}^{2}H^{-1})^{1/3}$, which is the length scale at which gravity is modified, is of the same order of the Hubble radius and we expect the gravity becomes 5-dimensional.  Nevertheless, it has been pointed out by Viaggiu~\citep{Viaggiu} that the condition (\ref{holobound}) is derived considering the Schwarzschild solution, which represented an exact solution for symmetric perturbation of Minkowskian spacetime. But when we apply this condition to Friedmann universes filled with dark energy leads to wrong results, due to avoidance of black hole formation when an small but finite cosmological constant is present. In the case of DGP models, modifications introduced in the metric of a spherically symmetric, static matter source, has been found in~\citep{Dai}. In this work the de-Sitter background is considered.  The main result lies in the absence of Birkhoff´s theorem for DGP theory, which means that for even spherically symmetric sources the exact distribution of matter affects the gravitational force external to the source.  These above results indicates that a fully consistent approach of the holographic dark energy in these models of modified gravity is currently under construction.  Considering this situation and despite the improvements realized in~\citep{Gong}, in order to consider higher dimensions in the holographic cutoff, we assume (\ref{holobound}) valid as a first approximation.\\

Nevertheless, the above discussion brings forward the following question: how differs both approaches for the holographic Ricci like dark energy? In what follows we evaluate numerically these differences. From equation (\ref{holodensity}) and using the expression for the desacceleration parameter $\displaystyle 1+q(z)=-\frac{\dot{H}}{H^2}$ we obtain:
\begin{eqnarray}
\rho(z) 	& = & 3\alpha \left[1-\frac{\beta}{\alpha}\left( 1+q(z) \right)\right]H^2(z)\label{fix1}\\
		& = & 3c^2(z)H^2(z)
\end{eqnarray}
where we have made the identification
\begin{eqnarray}
c^2(z) & = & \alpha\left[1-\frac{\beta}{\alpha}\left( 1+q(z) \right)\right]
\label{c2Function}
\end{eqnarray}
This means that the Ricci cut-off can be seen as a generalization of the holographic dark energy density $\rho=3c^2 H^2$, with $c=cte$. This condition has already been discussed in~\citep{Pavon}, where is pointed out that if $c^2(z)$ grows with time the bound given by the holographic condition progressively saturate up to full saturation when, asymptotically $c^(z)$ becomes constant.\\
If we rewrite equation (\ref{fix1}) as :
\begin{eqnarray}
\rho(z) & = & 3\bar{c}^2(z)H(z),\label{fix2}
\end{eqnarray}
with $\bar{c}^2(z)=c^2(z)H(z)$ the holographic Ricci bound mimics the holographic bound in~\citep{Gong} for a five dimensional gravity with $H$ as IR cutoff and a variable $\bar{c}^2(z)$ parameter. We will evaluate the variations of $\bar{c}^2(z)$ in the range of $0\leq z\leq 1$. Using the expression for $q(z)$ in terms of $H$ and $\dot{H}$ we can write
\begin{eqnarray}
q(z) & = & \frac{(1+z)}{E(z)}\frac{d E(z)}{dz}-1 \label{desaccelerationequation}
\end{eqnarray}
where $E(z)$ defined in Section \ref{SectionHubbleParameter} comes from the solution of the differential equation (\ref{Eq-HubbleODE-Redshift}), using the information of this section and the parameters obtained with the data analysis we obtain

\begin{figure}[h*]
\begin{center}
\includegraphics[width=8.5cm]{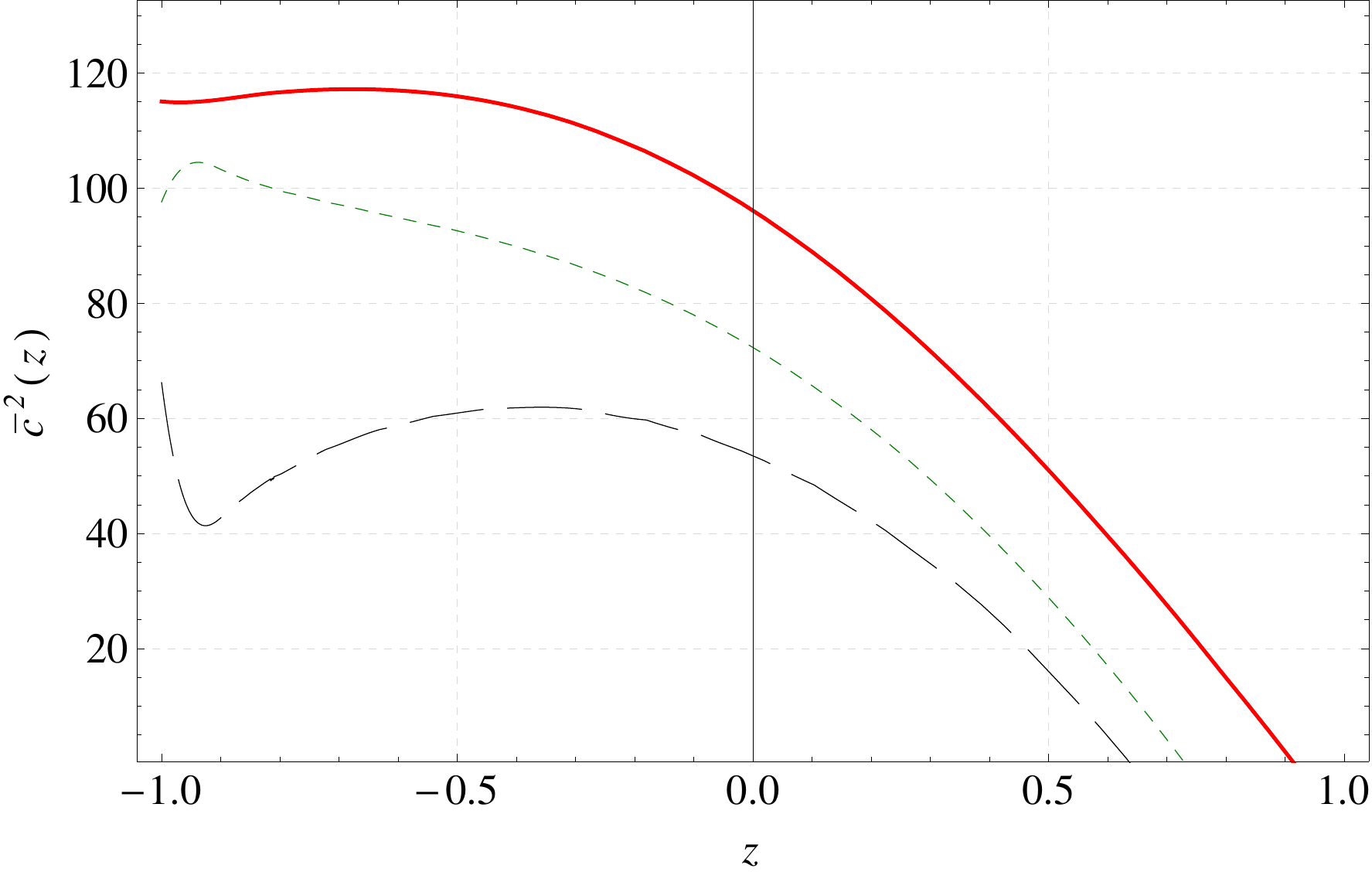}
\caption{Evolution of the function $\bar{c}^2(z)$ defined from equation (\ref{fix2}), in the late-time and future of the Universe. The red solid line corresponds to the best-fit values $(r_cH_0=1.3, \alpha=2.1, \beta=2.45, \Omega_{\rm m0}=0.46)$, while the green short dashed line and black long dashed line correspond to the arbitrary values of $(3, 1.1, 3, 0.3)$ and $(6, 4, 8, 0.4)$ respectively, with the purpose to illustrate the behavior of $\bar{c}^2(z)$ for other values. 
}
\label{FIG1}
\end{center}
\end{figure}
From Figure \ref{FIG1} we can see directly that the behavior of $\bar{c}^2(z)$ for future times its almost constant for the best-fit parameters (red solid line). This evaluation indicates that our holographic energy density is proportional to the Hubble parameter for future times, which is similar to the behavior of the model discussed in~\citep{Gong}.

\section{ Discussion and conclusions}

We have investigated the viability of a cosmological model composed by a Ricci-like holographic dark energy in the non  self-accelerated branch of a DGP braneworld, to explain the late-time accelerated expansion of the Universe.

We have discussed the holographic approach in higher dimensional gravity, indicating that it is required further research in order to have a fully consistent formulation. Nevertheless, an improvement in which the black hole solution in five dimension is taken into account, was compared with our approach, showing that for future late times both holographic densities behaves in similar way.

For the dark energy domination phase in the normal branch $\epsilon=-1$ , the model presents a de Sitter like expansion, and the conditions for the parameter are $\alpha>1$ and $\displaystyle \beta>-\left(1-\alpha+\frac{1}{r_c H_0}\right)$.

When we computed in the positive branch the confidence interval for $r_cH_0$, we found that the leakage length scale  is well compatible with positive values. Given that $r_c$ is a length scale, the minimum requirement of positive values for this parameter is satisfied. For the case when we consider the negative branch $\epsilon = -1$ of the DGP braneworld, we find that the parameters are constrained to be $1.2^{+3.5}_{-0.2}$, $\alpha=2.1^{+3.4}_{-1.1}$ and $\beta=2.45^{+5.5}_{-1.4}$ (see Table \ref{TableBestEstimated} (ii)), satisfying therefore the conditions for WEC and for the dark energy domination phase. So, the data indicates that the DGP model with an holographic Ricci-like dark energy might be a valid cosmological model.

On the other hand, in order to assess the viability of the present Ricci-like holographic DGP model to explain  the late-time accelerated expansion of the Universe, we compare the flat $\Lambda$CDM model with our present model by using the Bayesian Information Criterion (BIC) \citep{BIC-Schwarz} to determine which model is the most favored by the observations, taking into account the number of free parameters and the minimum value of $\chi^2$. 

In general, the value of BIC for Gaussian errors of the data used, is defined as
\begin{equation}
\text{BIC} = \chi^2_{\rm min} + \nu \ln N,
\end{equation}
where $\nu$ and $N$ are the  number of free parameters of the model and the number of data used respectively. The model favored by the observations compared to the others corresponds to that with the smallest value of BIC. In general, a difference of 2 in BIC between two models is considered as an evidence against the model with the higher BIC, and a difference of 6 is a strong evidence.

In  the last column of table \ref{TableBestEstimated} are shown the BIC values for holographic DGP and flat $\Lambda$CDM models.
We find that the flat $\Lambda$CDM model is  the favored model by the data according to BIC, compared with the holographic DGP model investigated in the present work.


		\section{acknowledgements}

NC and SL acknowledge the hospitality of the Physics Department of
Universidad de La Frontera where part of this work was done. SL and
FP acknowledge the hospitality of the Physics Department of
Universidad de Santiago de Chile. We acknowledge the support to this
research by CONICYT through grant 1110076 (SL) and grant 1140238 (NC). This work was
also supported from DIUFRO DI14-0007, Direcci\'{o}n de
Investigaci\'{o}n y Desarrollo, Universidad de La Frontera (FP) and VRIEA-DI-PUCV
 grant 037.377/2014, Pontificia Universidad Cat\'olica de Valpara\'iso (SL).
A.A. acknowledges the Mexico-Harvard Fellowship, the NSF for AST12-11196 and the Instituto Avanzado de Cosmolog\'ia of Mexico.
We acknowledge the use of \textit{emcee: The MCMC Hammer} Foreman-Mackey et al. (2013) \citep{emcee}, and  the \textit{triangle.py} python package \citep{TriangleCorner}.


\begin{thebibliography}{1}
\bibitem{Perlmutter} S. Perlmutter, et al., Astrophys. J. 517 (1999) 565;
P.M. Garnavich, et al., Astrophys. J. 493 (1998) L53; A.G. Riess, et al.,
Astron. J. 116 (1998) 1009; D.N. Spergel, et al., Astrophys. J. Suppl. 148
(2003) 175; A.G. Riess, Astrophys. J. 607 (2004) 665. [2] P.J.E. Peebles, B.
Ratra, Rev. Mod. Phys. 75

\bibitem{WMAP} Nine-Year Wilkinson Microwave Anisotropy Probe (WMAP) Observations: Cosmology Results Hinshaw, G.F., et.al.,2013, ApJS., 208, 19H.
\bibitem{DGP} G. R. Dvali, G. Gabadadze, and M.Porrati, Phys. Lett. B 485
(2000) 208.

\bibitem{Deffayet} C. Deffayet, Phys. Lett. B 502 (2001) 199.

\bibitem{Deffayet1} C. Deffayet, G.R. Dvali, G. Gabadadze, Phys. Rev. D 65
(2002) 044023.

\bibitem{Koyama} K. Koyama Gen. Rel. Grav. 40 (2008) 421.

\bibitem{Miao Li} M. Li, X. Li, S. Wang and Y. Wang, Commun. Theor. Phys.
56, 525 (2011) [arXiv:1103.5870].

\bibitem{Koyama1} K. Koyama, Class. Quant. Grav. 24, R231 (2007)
[arXiv:0709.2399 [hep-th]].

\bibitem{Lue} A. Lue and G. D. Starkman, Phys. Rev. D 70, 101501 (2004)
[arXiv:astro-ph/0408246]; V. Sahni and Y. Shtanov, JCAP 0311, 014 (2003)
[arXiv:astro-ph/0202346].

\bibitem{Lazkoz} R. Lazkoz, R. Maartens and E. Majerotto, Phys. Rev. D 74,
083510 (2006) [arXiv:astro-ph/0605701].

\bibitem{Mariam} M. Bouhmadi-Lopez, JCAP 0911:011 (2009)

\bibitem{MariamChimento} M. Bouhmadi-Lopez and L. Chimento,
Phys.Rev.D82:103506 (2010).

\bibitem{Cohen} A. G. Cohen, D. B. Kaplan and A. E. Nelson, Phys. Rev. Lett.
82, 4971 (1999) [hep-th/9803132].

\bibitem{Hsu} S. D. H. Hsu, Phys. Lett. B 594, 13 (2004) [hep-th/0403052].

\bibitem{Li} M. Li, Phys. Lett. B 603, 1 (2004) [hep-th/0403127].

\bibitem{Gonzalez} P.F. Gonz\'alez-D\'iaz, Phys. Rev. D27 (1983) 3042; G. t
Hooft, gr-qc/9310026; L. Susskind, J. Math. Phys. 36, 6377 (1995)
[hep-th/9409089].

\bibitem{Saridakis} E. N. Saridakis, JCAP 04, 020 (2008).

\bibitem{Wu} X. Wu, R. Cai, Z. Zhu, Phys.Rev.D77:043502 (2008)
arXiv:0712.3604 [astro-ph]

\bibitem{Gong} Y. Gong and T. Li, Phys. Lett. B 683, 241 (2010).

\bibitem{Liu} D. Liu, H. Wang, and B. Yang, Phys.Lett. B694 (2010) 6-9,
arXiv:1009.3776 [astro-ph.CO].

\bibitem{Mariam1} M. Bouhmadi-Lopez, A. Errahmani and T. Ouali, Phys. Rev. D
84, 083508 (2011)

\bibitem{Mariam2} M. H. Belkacemi, M. Bouhmadi-Lopez, A. Errahmani and T.
Ouali, arXiv:1112.5836 [gr-qc].

\bibitem{CGao} C. Gao, X. Chen and Y. Shen, Phys.Rev. D79 (2009) 043511. Kazuharu Bamba · Salvatore Capozziello ·
Shin’ichi Nojiri · Sergei D. Odintsov Astrophys Space Sci (2012) 342:155–228.

\bibitem{Oliveros} L. N. Granda and A. Oliveros, Phys. Lett. B669 (2008) 275-277

\bibitem{Yuting} Y. Wang, and L. Xu, Phys.Rev.
D81 (2010) 083523.

\bibitem{Feng} B. Feng, X. L. Wang and X. M. Zhang, Phys. Lett. B 607, 35 (2005);
B. Feng, M. Li, Y. S. Piao and X. M. Zhang, Phys. Lett. B 634, 101
(2006).

\bibitem{Samuel} S. Lepe and F. Pe\nn a, Eur. Phys. J. C (2010) 69:575-579.

\bibitem{Yu} F. Yu and J. Zhang, arXiv:1305.2792 [astro-ph.CO].

\bibitem{Samuel1} F. Ar\'evalo. P. Cifuentes, S. Lepe and F. Pe\nn a, Astrophys.Space Sci. 352 (2014) 899-907.

\bibitem{SWang} S.Wang, Y.Wang, and X. Zhang, JCAP 0912, 014 (2009). [arXiv:0910.3855]; M.
Li, X. Li, S.Wang, and X. Zhang, JCAP 0906, 036 (2009). [arXiv:0904.0928];
X. Zhang, [gr-qc/0909.4940]; L. Xu, [astro-ph.CO/0907.1709]; H. Wei,
[gr-qc/0902.2030]; S. Wu, P. Zhang, and G. Yang, [astro-ph/0809.1503].

\bibitem{PlanckCollaboration}
Planck Collaboration, P. A. R. Ade, et al., arXiv:1303.5076,2013.

\bibitem{Betoule2014}  Betoule et al. 2014, [arXiv:1401.4064].

\bibitem{TrippEquation} Tripp 1998, A\&A, 331, 815.

\bibitem{Avelino} A. Avelino and U. Nucamendi JCAP04(2009)006

\bibitem{Farooq} Omer Farooq et al. 2013 ApJ 764 138

\bibitem{Busca} Busca Nicolas \textit{et al.} 2012, [arXiv:1211.2616]

\bibitem{Blake1} Blake C. et al. 2011, MNRAS, \textbf{418}, 1725

\bibitem{HzData} Blake C. et al. 2012, MNRAS, \textbf{425}, 405; Chuang,
Chia-Hsun, Yun Wang 2012, MNRAS, \textbf{426}, 226; Reid, B.A., L. Samushia,
M.White et al. 2012, [arXiv:1203.6641]; Xu, X., Cuesta, A.J., N. Padmanabhan
et al. 2012, [arXiv:1206.6732]

\bibitem{RiessHzToday} A. G. Riess, L. Macri, S. Casertano, H. Lampeitl, H. C.
Ferguson, A.V. Filippenko, S.W. Jha, W. Li, and R.
Chornock, Astrophys. J. 730, 119 (2011); 
A.G. Riess, L.
Macri, S. Casertano, H. Lampeit, H. C. Ferguson, A.V.
Filippenko, S.W. Jha, W. Li, R. Chornock, and J. M.
Silverman, Astrophys. J. 732, 129(E) (2011).

\bibitem{SNojiri} S. Nojiri, S. Odintsov and S. Tsujikawa Phys. Rev. D 71,
063004 (2005) [hep-th/0501025]

\bibitem{BIC-Schwarz} G. Schwarz, Ann. Stat., 6, 461 (1978). 

\bibitem{Viaggiu} S. Viaggiu, Mod.Phys.Lett. A29 (2014); arXiv: 1312.2889 (2013)

\bibitem{Dai} D. Dai, I. Maor and G. D. Starkman, Phys.Rev. D77 (2008) 064016.  

\bibitem{Pavon} D.Pavon and W. Zimdahl, Phys. Lett. B628 (2005) 206-210.

\bibitem{AffineInvariantMCMC} Jonathan Goodman and Jonathan Weare, Communications in Applied Mathematics and Computational Science, Vol \textbf{5} (2010), No. 1, 65-80. DOI: 10.2140/camcos.2010.5.65

\bibitem{emcee} Daniel Foreman-Mackey, David W. Hogg, Dustin Lang and Jonathan Goodman, 
Publications of the Astronomical Society of the Pacific, Vol. 125, No. 925 (March 2013) (pp. 306-312)

\bibitem{TriangleCorner} Dan Foreman-Mackey ; Adrian Price-Whelan ; Geoffrey Ryan ; Emily ; Michael Smith ; Kyle Barbary ; David W. Hogg ; Brendon J. Brewer, http://dx.doi.org/10.5281/zenodo.10598

\end{thebibliography}
\end{document}